\documentclass[10pt,a4paper]{article}
\usepackage{qsl-template}
\usepackage{mathrsfs} 
\usepackage[T1]{fontenc} 
\usepackage{lmodern} 
\usepackage{amsmath}
\usepackage{graphicx}
\usepackage{tikz}
\usepackage{etoolbox}
\usepackage{ragged2e}
\usepackage{subcaption}

\title{\fontseries{b}\selectfont Dispersion in nonlinear interferometry: implications for optical coherence tomography with undetected photons}

\AuthorsBlock
{
  \Author{Ivan Zorin}{1\orcid{0000-0002-2089-5716} },
  \Author{Paul Gattinger}{1\orcid{0000-0002-0120-9100}},
}
{
  \Affil{1}{Research Center for Non-Destructive Testing, Linz, Austria}
}
{E-mail: \fontseries{b}\selectfont ivan.zorin@recendt.at}

\begin{document}

\maketitle

\abstract{
Nonlinear SU(1,1) quantum interferometers based on non-degenerate optical parametric down-conversion exhibit strong unbalanced group velocity dispersion (GVD). This feature is intrinsic to this type of interferometer as correlated photons of vastly different frequencies propagate through a dispersive nonlinear crystal; consequently, the dispersion arises from the source itself. The resulting GVD degrades the axial point-spread function (PSF) in optical coherence tomography (OCT) with undetected photons; and physical compensation is less straightforward, in particular for non-degenerate broadband regimes due to the limited number of suitable materials.
In this contribution, we analyze dispersion in bulk nonlinear interferometry and describe its implications for OCT imaging. Aspects of hardware compensation are addressed, and a novel empirical numerical method of compensation is proposed. The approach is based on the extraction of the phase component directly from the time-domain modality (high precision linearized quantum Fourier transform infrared spectrometer) and its injection into the mid-IR spectral-domain OCT signals (central wavelength of around 3770~nm) before the Fourier transform. The proposed method is compared with an alternative numerical technique. The results demonstrate a 2.2-fold improvement in axial resolution and outperform the alternative correction method in overall imaging performance.
}

\section{Introduction}

Nonlinear SU(1,1)\footnote{(SU) - (Special Unitary) group} interferometry, rooted in fundamental quantum phenomena such as entanglement and bi-photon interference~\cite{Zou1991,Mandel1991}, has emerged as a promising alternative to state-of-the-art classical optical metrology~\cite{Chekhova:16,10.1063/5.0004873}, enabling a range of quantum-based sensing modalities, including sensing with undetected photons~\cite{Lemos_2014, BarretoLemos:22}.
These techniques rely on non-degenerate spontaneous (in the low-gain regime) parametric down-conversion (SPDC), which enables the generation of correlated photon pairs across various spectral ranges. In this process, a pump of a higher energy is converted into two photons of lower frequencies: signal and idler. Thus, an SPDC source, as a core component of the SU(1,1) interferometer, provides the necessary degrees of freedom to access and operate in domains challenging for classical instrumentation. 
Sensing with undetected photons facilitates access to information in the spectral range of interest (undetected probing idler photons) by means of detection of the interference of correlated signal photons. This eliminates the need for direct optics and detectors in the challenging spectral range, enabling the use of high-sensitivity and mature shot-noise limited detection technologies.

Since metrology with undetected photons is inherently interferometric, it was naturally adapted and applied to well-established classical methods, such as coherent imaging and microscopy~\cite{kviatkovsky_microscopy_2020,325251662fcd4d999a1d597579d67827,placke_mid-ir_2026}, Fourier-transform infrared (IR) spectroscopy~\cite{Lindner2020,Lindner:21,Lindner2022,Mukai2022,Lindner2023,Tashima2024,PhysRevApplied.22.044015,Gattinger:25} (quantum FTIR, QFTIR, in this case), optical coherence tomography (OCT) and low-coherence interferometry ~\cite{Valles2018,Paterova2018,10.1063/5.0016259,Vanselow:20,https://doi.org/10.1002/qute.202300299,Zotti:25}. In addition, several unconventional interferometric approaches to spectroscopy have been demonstrated due to the specific features and peculiarities of SU(1,1) schemes~\cite{Kalashnikov2016,Paterova2022,Kaufmann2022,Cardoso2024,10.1063/5.0242197}.
Most promising sensing applications of nonlinear interferometry are reported in the mid-IR spectral range\textemdash a spectral domain characterized by constrained technological performance and high costs of classical approaches. Thus, advantages over traditional instruments are easier to achieve\footnote{There is a potential of extension to even longer terahertz waves~\cite{https://doi.org/10.1002/qute.202500271}}. It should be noted that most of the publications have appeared within the last five years, emphasizing the emerging nature and the continued advancement of this metrological field.

Classical mid-IR OCT is a relatively recent technique, first experimentally demonstrated in 2018, and has shown clear advantages for non-destructive testing~\cite{su_perspectives_2014,zorin_mid-infrared_2018,israelsen_real-time_2019,zorin_mid-infrared_2022,yagi_mid-infrared_2024}; however, the demonstrated solutions relied on complex or noisy detection schemes and light sources (e.g., supercontinuum sources) that alone are significantly more expensive than an entire quantum system including an SPDC source and a pump laser.
By contrast, mid-IR OCT based on nonlinear interferometry~\cite{Vanselow:20} is one of the most promising recent alternative approaches, as it offers competitive sensitivity and performance at costs attractive for applied research.

OCT implemented in the nonlinear interferometric scheme possesses distinct physical features that fundamentally distinguish it from classical OCT, despite the strong resemblance at first glance. Non-degeneracy of the SPDC emission implies that the interference pattern detected in the signal domain (i.e., near-IR) is dictated by the joint interplay of all involved photons of different frequencies. Hence, strong unmatched dispersion arises in the crystal itself, and is thus an intrinsic property of the sources (unless high-order nonlinear dispersion is canceled at the design stage in the waveguide architecture~\cite{Roeder_2024}). Another consequence is that hardware dispersion compensation is sophisticated, particularly for strong non-degeneracy and broadband emission. Since the reference and sample arms contain different photon frequencies, simple matching of the materials in the interferometric arms is not feasible.

In this contribution, we introduce, describe, and analyze the dispersion in nonlinear interferometry and its implications on OCT imaging with undetected photons. We investigate dispersion-related peculiarities arising from the propagation of correlated signal and idler photons in a medium and their effects on the interferometric signals.
In the experimental section, we propose a novel numerical method for higher-order dispersion compensation that leverages the relation between the mutual bi-photon coherence function (time-domain) and the complex spectrum (frequency-domain) as described by the Wiener–Khinchin theorem and its cross-correlation generalization~\cite{Mandel_Wolf_1995}.
Since FTIR and OCT are, in essence, the same system, while the measured quantities are interchanged, we employ the QFTIR modality built upon the same interferometer (used for OCT) to directly measure the phase (in the style of dispersive spectroscopy) and compensate for it in the OCT imaging modality. The derived phase can then be stored for any sequentially sampled spectral interferograms. The novel method is compared with other numerical methods conventionally used in classical OCT.

\section{Dispersion in nonlinear interferometry}

\begin{figure}[hbt]
    \begin{subfigure}[b]{0.5\linewidth}
        \centering
        \begin{tikzpicture}
            \node[anchor=south west, inner sep=0] (img) at (0,0) {%
                \includegraphics[
                    trim=40mm 10mm 140mm 15mm,
                    clip,
                    width=\textwidth
                ]{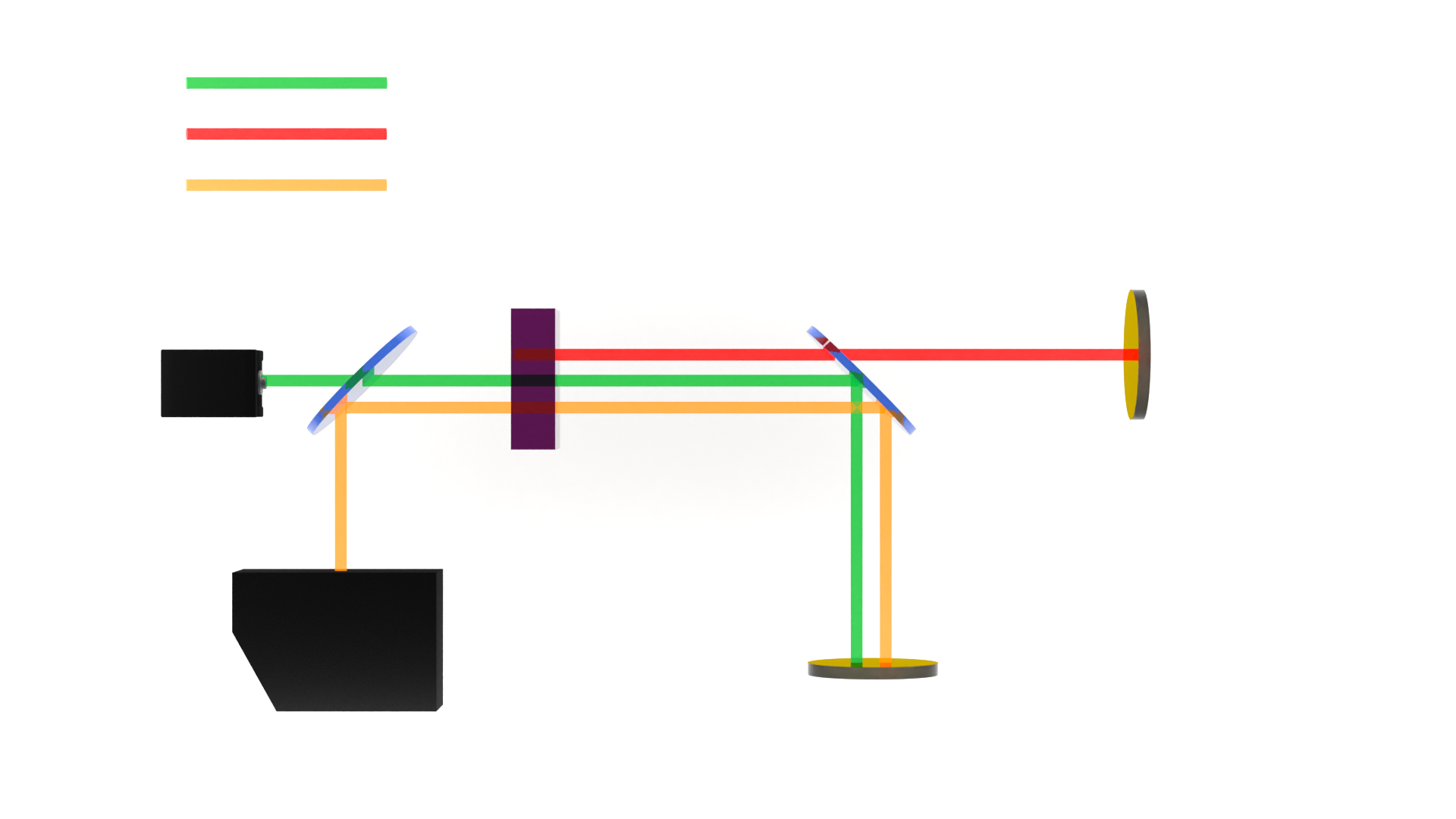}%
            };
            \begin{scope}[x={(img.south east)}, y={(img.north west)}]
            
             \draw (0.12,0.63) node[]{\color{black}\footnotesize Pump};
             \draw (0.42,0.4) node[]{\color{black}\footnotesize $\chi^{(2)}$};
             \draw (0.7,0.66) node[]{\color{black}\footnotesize DM1};
             \draw (0.29,0.66) node[]{\color{black}\footnotesize DM2};
             \draw (0.97,0.7) node[]{\color{black}\footnotesize M1};
             \draw (0.74,0.12) node[]{\color{black}\footnotesize M2};
            \draw (0.25,0.07) node[]{\color{black}\footnotesize Spectrometer};
            \draw (0.06,0.93) node[]{\color{black}\footnotesize $\omega_p$};
            \draw (0.06,0.86) node[]{\color{black}\footnotesize $\omega_i$};
            \draw (0.06,0.79) node[]{\color{black}\footnotesize $\omega_s$};
            \end{scope}
        \end{tikzpicture}
        \caption{Conceptual arrangement of a SU(1,1) Michelson type interferometer}\label{fig:su11b}
    \end{subfigure}
    \hfill
    \begin{subfigure}[b]{0.475\linewidth}
        \centering
        \includegraphics[width=\textwidth]{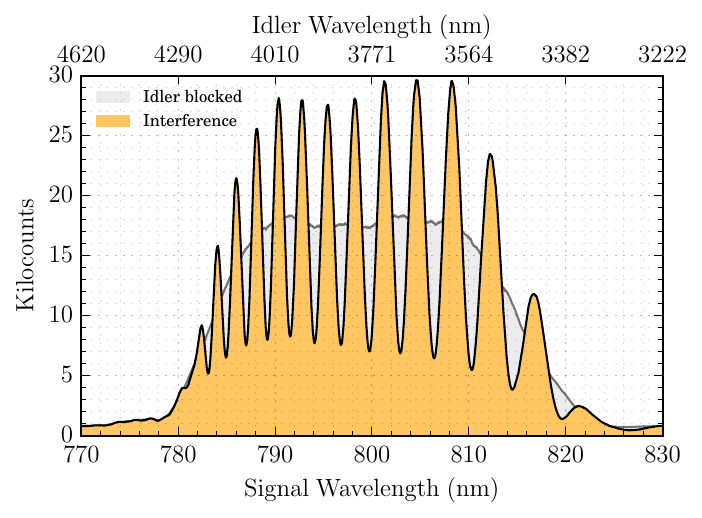}
        \caption{Quantum interference recorded in the signal domain $\lambda_s = 2\pi c/{\omega_s}$ (system detailed in the next section)}
        \label{fig:fringes}
    \end{subfigure}
    \caption{ Principles of nonlinear interferometry and sensing with undetected photons: (a) Simplified schematic of a nonlinear interferometer (base OCT unit) employing a non-degenerate SPDC source (a $\chi^{(2)}$ crystal pumped by a monochromatic laser). DM1 forms the reference (signal and pump) and sample (idler) arms. Interference is detected in the signal channel (signal photons are reflected by DM2) by a grating spectrometer. The interference pattern is governed by the amplitude and phase response experienced by the idler photons. (b) Spectral interferogram measured in the frequency domain.}
    \label{fig:nonlinear_interferometer}
\end{figure}
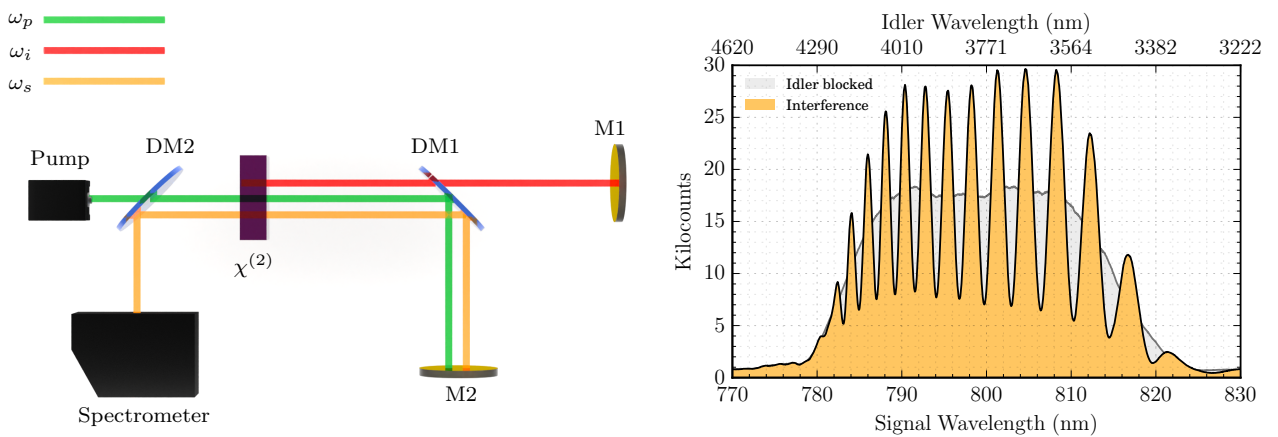

Nonlinear interferometers\textemdash also known as SU(1,1) or bi-photon interferometers\textemdash are quantum-optical systems that employ parametric nonlinear processes ($\chi^{(2)}$ or $\chi^{(3)}$) for amplitude- and phase-sensitive metrology. 
In the principal structural schemes of these interferometers, beamsplitters are replaced with active nonlinear parametric amplifiers. The systems can be realized in the low- and high-gain operational regimes~\cite{PhysRevA.33.4033,Chekhova:16}.
A variety of nonlinear interferometer architectures have been demonstrated. In this work, we focus on a Michelson-type arrangement with a non-degenerate SPDC source in the low-gain regime, as it is a common and straightforward configuration for applied photonic quantum sensing, well-suited for practical OCT with undetected photons~\cite{PhysRevA.106.033702}. The conceptual optical scheme of such a bi-photon interferometer along with a description of its core principles is illustrated in Fig.~\ref{fig:nonlinear_interferometer}.

The core element of the nonlinear interferometer displayed in Fig.\ref{fig:su11b} is a non-degenerate broadband SPDC source of energy- and momentum-entangled photons. In essence, the SPDC source is a $\chi^{(2)}$ nonlinear crystal pumped by a monochromatic laser at the frequency $\omega_p$. In the SPDC process, correlated signal and idler photons at frequencies $\omega_s$ and $\omega_i$ are generated; energy conservation imposes the strict frequency correlation:
\begin{equation}
\omega_p=\omega_s+\omega_i.
\label{freq-corr}
\end{equation}

In order to tailor nonlinear interferometry for such a low-coherence imaging modality as OCT, group velocity matching is employed to generate ultra-broadband, substantially non-degenerate photon pairs~\cite{Vanselow:19}. 
Similar to the classical electric field amplitude in conventional OCT, the broadband nature can be expressed as the probability amplitude $f(\omega_s)$ of photon emission in a certain $\omega_s$ mode dictated by the phase-matching conditions in the crystal. 

The interferometer is formed by a dichroic mirror (DM1), which plays a role analogous to a classical beam splitter: it reflects the pump ($\omega_p$) and signal ($\omega_s$) photons into one arm and transmits the idler photons ($\omega_i$) into the idler (sample) arm~\textemdash according to the convention used for probing. After the second pass through the crystal, which may trigger a second SPDC process, interference can be observed in the signal channel (the signal photons are selected by DM2), as dictated by the principles of path-indistinguishability. 
In general, interference can be seen in both domains, however, metrology with undetected photons implies detection only in one~\textendash commonly the signal domain. 

Figure~\ref{fig:fringes} serves to illustrate signal and corresponding idler domains along with raw spectral interferograms; the experimental data are obtained from a mid-IR nonlinear interferometer based on a periodically poled potassium titanyl phosphate (ppKTP) crystal (poling period $\Lambda$=20.45~\textmu m). In contrast to classical implementations, OCT based on nonlinear interferometry yields a constant total number of detected signal photons (the integrals of the orange and gray areas in Fig.~\ref{fig:fringes} are equal). Instead, the visibility magnitude is determined by the degree of indistinguishability of the signal and idler photon paths~\cite{Mandel1991}. The interference pattern vanishes if the idler photons are absorbed or scattered~\textendash i.e., corresponding signal photons become in principle distinguishable. Hence, the information in the idler domain can be accessed using detection in the spectral range of the signal photons. The phase experienced by the idler and signal can be measured in a similar manner~\cite{PhysRevA.33.4033,BarretoLemos:22}, albeit with consideration of the inherent peculiarities of SU(1,1) interferometers.

In nonlinear interferometers that form the base for OCT with undetected photons, the phase that modulates the intensity distribution is the collective phase accumulated by the involved photons~\cite{Chekhova:16}:
\begin{equation}
\Delta\varphi(\omega_s, \omega_p)=\varphi_p(\omega_p)-\varphi_s(\omega_s)-\varphi_i(\omega_i).
\label{phi-g}
\end{equation}

This consolidated phase term complicates the description of dispersion in nonlinear interferometry and limits the effectiveness of hardware-based compensation. In particular, even an empty nonlinear interferometer (without e.g. probing optics) can be intrinsically unbalanced and exhibit uncompensated dispersion, since photons of different frequencies are generated and propagate within the nonlinear crystal itself. Although the interferometer operates with non-classical light, an analysis of dispersion can be performed by applying the classical mathematical framework developed for conventional OCT.

The phase experienced by a photon during the propagation is defined as:
\begin{equation}
\varphi(\omega)=k(\omega)\,z,
\end{equation}
where $\omega$ is the angular frequency, $z$ is the propagation length (e.g., position of the reflector or propagation through any material) and $k(\omega)$ is the effective propagation constant defined as:
\begin{equation}
k(\omega)=\frac{n(\omega)\,\omega}{c};
\end{equation}

$n(\omega)$ is the frequency-dependent refractive index of the medium and $c$ is the speed of light.

Thus, each of the phase components involved in Eq.\ref{phi-g} has the form of $\varphi_\mu(\omega)=k_\mu(\omega)z_\mu$ where $\mu=p,s,i$ (pump, signal, idler respectively) with $z_\mu$ displaying the propagation of the corresponding photon in the crystal.

Expanding $k(\omega)$ about a central frequency $\omega_0$, as commonly done to express group velocity dispersion (and higher order contributions) in classical OCT, yields\cite{drexler_optical_2015}:
\begin{equation}
k(\omega)\approx k(\omega_0)+k^{(1)}(\omega_0)(\omega-\omega_0)+\frac{1}{2}k^{(2)}(\omega_0)(\omega-\omega_0)^2+\cdots,
\end{equation}
with coefficients defined as derivatives:
\begin{equation}
k^{(m)}=\left.\frac{\partial^m k}{\partial\omega^m}\right|_{\omega_0}, \qquad m=0,1,2,\ldots
\end{equation}

Thus, the phase can be rewritten as:
\begin{equation}
\varphi(\omega)\approx z\left[k(\omega_0)+k^{(1)}(\omega-\omega_0)+\frac{1}{2}k^{(2)}(\omega-\omega_0)^2+\cdots\right],
\label{phi-ext}
\end{equation}
depicting also dependence of the phase (linear and high-order nonlinear) on $z$.


Let $\omega_{s_0}$ be the central signal frequency and $\omega_{i_0}=\omega_p-\omega_{s_0}$ the corresponding idler frequency.
Relying on the quantum principles of the frequency correlation (Eq.~\ref{freq-corr}) it is possible to define:
\begin{equation}
\omega_s-\omega_{s_0}=-(\omega_i-\omega_{i_0})=-\Omega,
\end{equation}
in order to express the total dispersion.

The signal and idler propagation constants can therefore be consistently rewritten for $\Omega$:
\begin{equation}
k_s(\omega_s)\approx k_s^{(0)}(\omega_{s_0})-k_s^{(1)}\Omega+\frac{1}{2}k_s^{(2)}\Omega^2+\cdots,
\end{equation}
\begin{equation}
k_i(\omega_i)\approx k_i^{(0)}(\omega_{i_0})+k_i^{(1)}\Omega+\frac{1}{2}k_i^{(2)}\Omega^2+\cdots.
\end{equation}

It should be noted that the propagation constant for the pump is not expanded, since the pump is assumed to be quasi-monochromatic; its spectral bandwidth is sufficiently narrow that higher-order dispersion terms can be neglected without appreciable error.

Substituting these expansions into Eq.~\ref{phi-g} and using Eq.~\ref{phi-ext} yields:
\begin{align}
\Delta\varphi(\omega_s)\approx\;
&\underbrace{\left[k_p^{(0)}(\omega_{p_0})z_p-k_s^{(0)}(\omega_{s_0})z_s-k_i^{(0)}(\omega_{i_0})z_i\right]}_{D}
\nonumber\\
&-\underbrace{\left(k_i^{(1)}z_i-k_s^{(1)}z_s\right)}_{\Delta\tau}\,\Omega
-\frac{1}{2}\underbrace{\left(k_s^{(2)}z_s+k_i^{(2)}z_i\right)}_{\Gamma^{(2)}}\,\Omega^2
+\cdots .
\label{del_phi-fin}
\end{align}

The terms in Eq.~\ref{del_phi-fin} contain three terms that correspond to zero-, first-, and second-order dispersion, respectively. Thus, $D$ is the constant phase offset; $\Delta \tau$ defines the axial position (round trip group delay mismatch i.e. $z_i/v_{g,i} - z_s/v_{g,s}$, where $v_g$ is the group velocity) and corresponding frequency of the fringes; and $\Gamma^{(2)}$ represents the effective group-velocity-dispersion (GVD) contribution, which, together with higher order terms, degrades the axial resolution in OCT imaging~\cite{drexler_optical_2015}.


The modulated spectrum defined in \cite{Vanselow:20} can then be generalized to include the high-order GVD term, constraining the axial resolution:
\begin{equation}
I(\omega_s)
 = f^2(\omega_s)\left(2+
r\cos\!\left[
D
+ \Delta\tau\,(\omega_s-\omega_{s_0})
- \frac{1}{2}\Gamma^{(2)}\,(\omega_s-\omega_{s_0})^2
\right]\right),
\end{equation}

where $f(\omega_s)$ is the probability amplitude of photon emission in the $\omega_s$ mode and $r$ is the combined signal-idler reflectivity amplitude; the factor of 2 arises due to double forward and backward propagation through the crystal, causing SPDC at each pass.

\begin{figure} [hbt]
 \centering
        \includegraphics[width=0.9\textwidth]{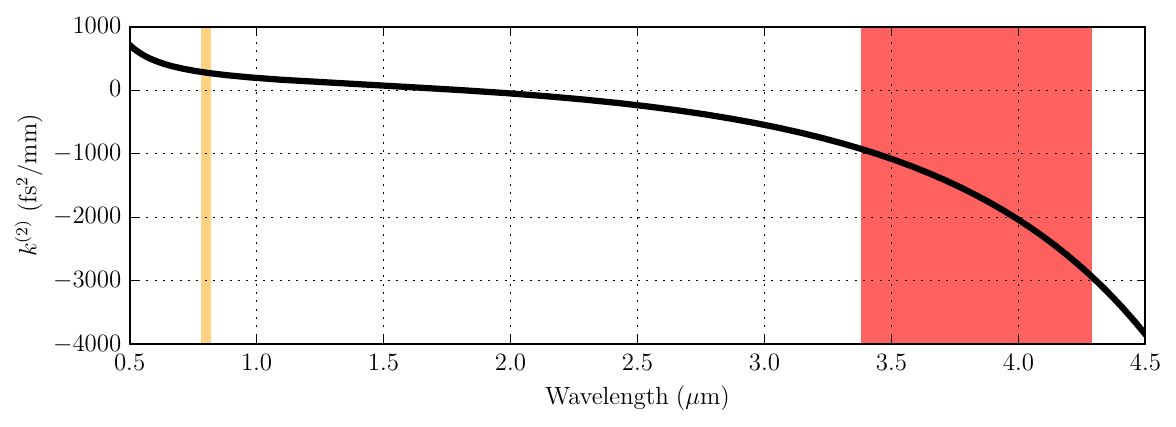}
 \caption{Group velocity dispersion of KTP crystal~\cite{10.1063/1.1375802} derived in the wavelengths $\lambda = 2\pi c/{\omega}$; signal and idler spectral ranges (660~nm pump, poling period of KTP $\Lambda$=20.45~\textmu m) are indicated in orange and red respectively; signal and idler have equal bandwidth in frequency domain (determined by the energy conservation rule $\omega_p=\omega_s+\omega_i$) and scale nonlinearly when displayed in wavelengths.}
\label{fig:gvd}
\end{figure}

Therefore, in contrast to classical OCT, where axial broadening is determined by the residual (differential) GVD mismatch between the reference and sample arms, OCT with undetected photons exhibits an effective second-order dispersion as the net sum of the signal and idler GVD contributions~\cite{Roeder_2024}. This behavior follows directly from the collective phase of interference
\(\varphi_p-\varphi_s-\varphi_i\) and the energy correlation
$\omega_p=\omega_i+\omega_s$ inherent to the nonlinear interferometer and generalized SPDC process, respectively. Thus, the effective dispersion that affects interferometric signals in SU(1,1) systems is the dispersion accumulated by a bi-photon during its propagation through the medium.
This point is illustrated in Fig.~\ref{fig:gvd} for $k^{(2)}$ of a KTP crystal, calculated using Sellmeier equations from~\cite{10.1063/1.1375802}. 

In the particular case of KTP, the crystal exhibits a zero-dispersion wavelength around 2~\textmu m, with normal (positive) dispersion in the signal region and anomalous (negative) dispersion in the idler spectral domain. A similar behavior is observed in lithium niobate. Since the GVD term $\Gamma^{(2)}$ contains the sum of the GVD contributions from both interacting photon bands, the total bi-photon group-delay dispersion accumulated in the crystal (inherent for an SPDC source) is partially self-compensated; however, the strong negative GVD within the idler band ultimately dominates. Interestingly, materials that possess positive dispersion across a broad range can be put in both arms to compensate for the dispersion as they effectively add a positive contribution either via $k_i^{(2)}$ or $k_s^{(2)}$ (when the sum is negative); although the absolute GVD and steepness, i.e. total dispersion should be considered as it might be low for the short wave signal domain: for instance, standard N-BK7 glass has a GVD of around 44.65~fs\textsuperscript{2}/mm at $\lambda$=800~nm~\cite{schott_optical_glass_datasheets_2026}. Potentially dense flint glasses can be an option for compensation in the signal arm (SF10 has 156.53~fs\textsuperscript{2}/mm GVD at $\lambda$=800~nm~\cite{schott_optical_glass_datasheets_2026}). 
Eq.~\ref{del_phi-fin} further indicates that, when the zero-dispersion wavelength lies between the signal and idler bands (as in KTP), a material exhibiting normal (positive) dispersion across both the signal and idler bands can be used to enhance dispersion compensation when placed in the combined pump-signal–idler beam; thus, it can be located after the crystal. These consequences are counterintuitive in the context of classical interferometry. 


 


\section{Dispersion compensation: spectral-domain mid-IR OCT with undetected photons}

\begin{figure}[hbt]
 \centering
\begin{tikzpicture}
  \node[anchor=south west,inner sep=0] (image) at (0,0,0) {\includegraphics[width=0.85\textwidth]{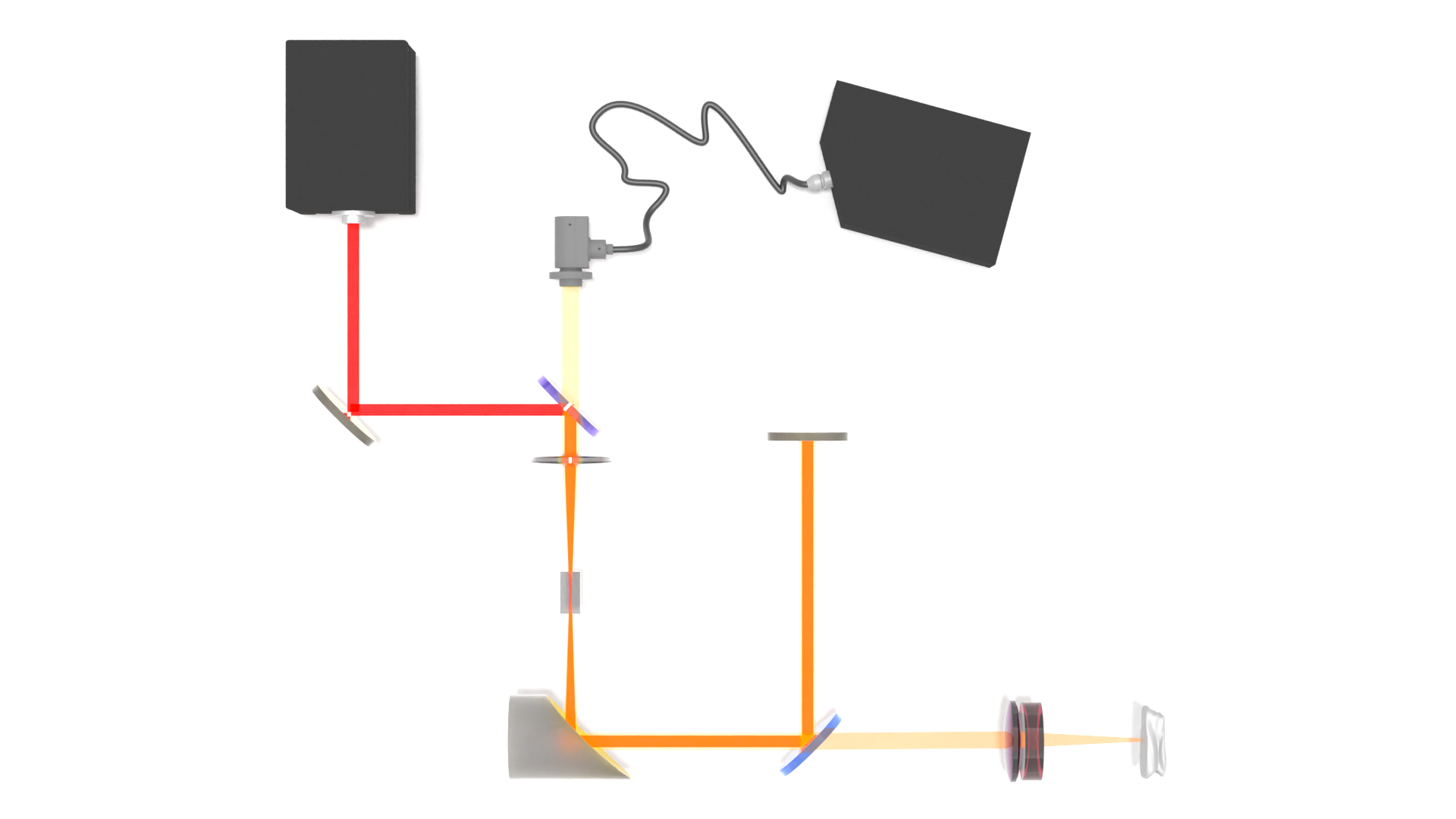}};
  \begin{scope}[x={(image.south east)},y={(image.north west)}]
    \draw (0.24,0.975) node[]{\color{black}\footnotesize Pump laser};
    \draw (0.64,0.93) node[]{\color{black}\footnotesize Grating spectrometer};
    \draw (0.42,0.275) node[rotate=90]{\color{black}\footnotesize $\chi^{(2)}$ crystal};
    \draw (0.33,0.1) node[rotate=-90]{\color{black}\footnotesize OPM};
    \draw (0.43,0.53) node[]{\color{black}\footnotesize CM};
    \draw (0.59,0.18) node[]{\color{black}\footnotesize DM};
    \draw (0.56,0.53) node[]{\color{black}\footnotesize Reference};
    \draw (0.56,0.5) node[]{\color{black}\footnotesize mirror};
    \draw (0.7,0.18) node[]{\color{black}\footnotesize Lens};
    \draw (0.79,0.18) node[]{\color{black}\footnotesize Sample};
    \draw [latex-latex,thick, color={black}] (0.59, 0.45) -- (0.59, 0.35);
    \draw (0.615,0.4) node[]{\color{black}\footnotesize $\delta$};
  \end{scope}
\end{tikzpicture}
\caption{\label{fig:setup} Spectral-domain OCT with undetected photons based on nonlinear low-gain SU(1,1) interferometer: 500 mW pump laser (660~nm), Lens - achromatic doublet (silicon and germanium) lens, $\chi^{(2)}$ crystal - ppKTP (periodically poled potassium titanyl phosphate, $\Lambda = 20.45$~\textmu m, $l=2.55$~mm), OPM - off-axis parabolic mirror, DM - dichroic mirror, CM - cold mirror.} 
\label{fig:system}
\end{figure}
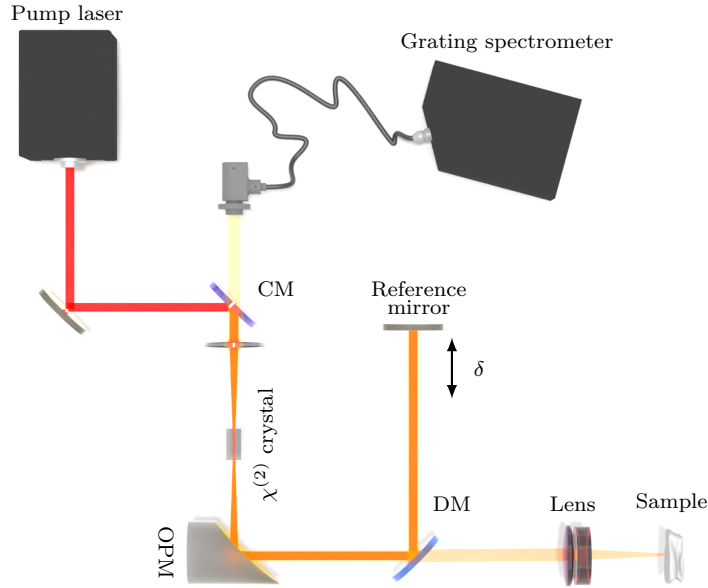

In order to demonstrate the effects of high-order dispersion and propose a tailored method for numerical compensation, an experimental mid-IR OCT system with undetected photons, schematically shown in Fig.~\ref{fig:system}, was employed. The setup is an actual arrangement of the conceptual scheme shown in Fig.~\ref{fig:nonlinear_interferometer}; a detailed description and characterization of the setup can be found in~\cite{Gattinger:25}.

The OCT system is implemented in the spectral-domain as a low-gain Michelson-type nonlinear interferometer with a double-pass of the pump laser (linearly polarized 660 nm, 500 mW, continuous wave) through a periodically poled potassium titanyl phosphate crystal (ppKTP, $\Lambda = 20.45$~\textmu m, $l=2.55$~mm).
The pump laser is injected into the system using a cold mirror (CM; high reflectivity at the pump wavelength), and focused by an achromatic lens into the ppKTP crystal to generate broadband signal–idler photon pairs.
An off-axis parabolic mirror (OPM) collimates the pump, signal, and idler photons. A custom dichroic mirror (DM, 5~mm calcium fluoride substrate) separates the photon paths into reference (pump-signal, reflected) and sample (idler, transmitted) arms, respectively. The pump–signal reference arm includes a reference mirror mounted on a motorized linear translation stage (omitted from Fig.~\ref{fig:system} for simplicity) employed for time-domain measurements of the QFTIR modality. The sample arm contains the probing optics, which focus the idler beam (approx. 60~pW) onto the sample.
After the second pass through the crystal, the signal photons possessing the bi-photon interference are directed through the CM and coupled into a high-resolution near-IR grating spectrometer (Ocean HR, Ocean Optics) using a single-mode fiber.
In the QFTIR modality, mirror scanning is tracked and linearized using a reference interferometer with a HeNe laser (632.8 nm); the acquired time-domain interferograms are resampled to ensure high-precision spatial sampling. For the time-domain measurements, the fiber is connected to a single point detector.

As the probing optics, an achromatic air-spaced doublet lens ($f=$50~mm, AC254-050-E, Thorlabs), consisting of silicon (2~mm) and germanium (2.668~mm), is used in the idler arm to focus the beam onto the sample. Since both silicon and germanium exhibit positive GVD in the mid-IR range~\cite{edwards_infrared_1980,levan_refractive_2016} (371.8~fs\textsuperscript{2}/mm for Si and 1105.2~fs\textsuperscript{2}/mm for Ge at the central wavelength), the doublet lens effectively compensates strong negative dispersion of the crystal (2.55~mm KTP, see Fig.~\ref{fig:gvd}) and the DM (-230.4~fs\textsuperscript{2}/mm~\cite{10.1063/1.555616}). 
This physical compensation yields a reduced positive net dispersion, the total group-delay dispersion changes from approximately -5501.5 to $+1883.1$~fs\textsuperscript{2} at the central wavelength.
Consequently, a noticeable reduction in the broadening of the time-domain interferograms and an improvement in axial resolution of OCT can be observed compared to the nonlinear interferometer with no additional optical elements.

In order to compensate the residual bi-photon GVD, we propose to use a second modality built into the very same nonlinear interferometer\textemdash QFTIR, reported in~\cite{Gattinger:25}. The QFTIR mode enables high-precision quantitative measurements of the amplitude and phase spectra; the latter contains the information necessary for the dispersion compensation.
In QFTIR the phase components can be neglected as in~\cite{Lindner:21} or used~\cite{Lindner:22} for dispersive spectroscopy. Hence, it is possible to exploit the precisely measured QFTIR phase to compensate for the bi-photon GVD and enhance the axial resolution of OCT with undetected photons.

Hereafter, we use wavenumber $\tilde{\nu}$ (reciprocal of wavelength) as the standard spectral unit, rather than angular frequency $\omega$, since it is suitable for both OCT (linear $k$-space) and IR spectroscopy (natural unit):

\begin{equation}
\tilde{\nu}=\frac{1}{\lambda}=\frac{\omega}{2\pi c}.
\end{equation}

\begin{figure}[hbt]
 \centering
        \includegraphics[width=0.8\textwidth]{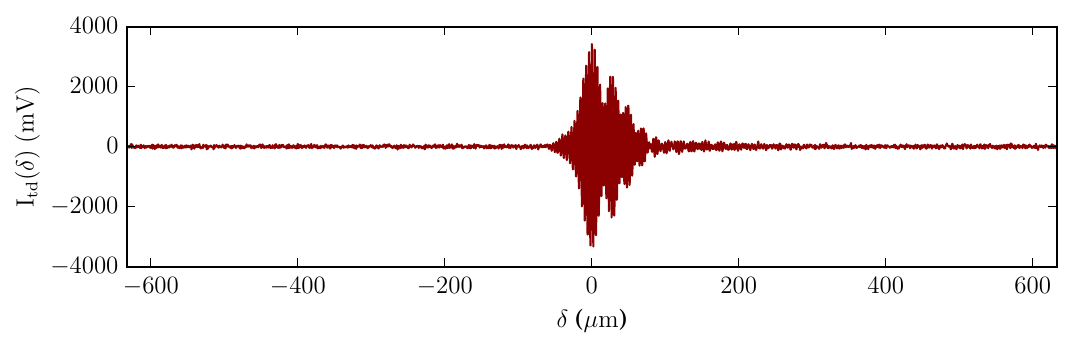}
 \caption{Time-domain interferogram recorded with a single point detector during simultaneous scanning of phases of signal and pump photons ($\varphi_p-\varphi_s$); raw signal of the QFTIR modality, contains amplitude as well as phase information.}
\label{fig:td-signal}
\end{figure}

In order to measure the net dispersion accumulated by the bi-photon, a time-domain real-valued interferogram $\mathrm{I_{td}}$ is recorded by scanning the optical path difference $\delta$ in the reference arm (in the style of time-domain OCT) using a single point detector. Instead of a sample, a single reflector that introduces no phase delays (gold mirror) is placed in the sample arm. The recorded time-domain signal is shown in Fig.~\ref{fig:td-signal}. The interferogram displays a central burst akin to classical FTIR~\cite{griffiths2007fourier}; however, it exhibits asymmetry due to the GVD and higher-order dispersion; similar signals can be experienced, for instance, in dispersive IR 
spectroscopy~\cite{birch_dispersive_1987,doi:10.1080/00107519008213783,Bell1967}. It should be noted that the measurements shown in Fig.~\ref{fig:td-signal} are resampled to a uniform spatial grid using the HeNe reference interferometer, with a sampling precision of $\lambda/2$ (HeNe).

\begin{figure}[htb]
 \centering
        \includegraphics[trim={2cm 3.1cm 0 2.0cm},clip,width=0.8\textwidth]{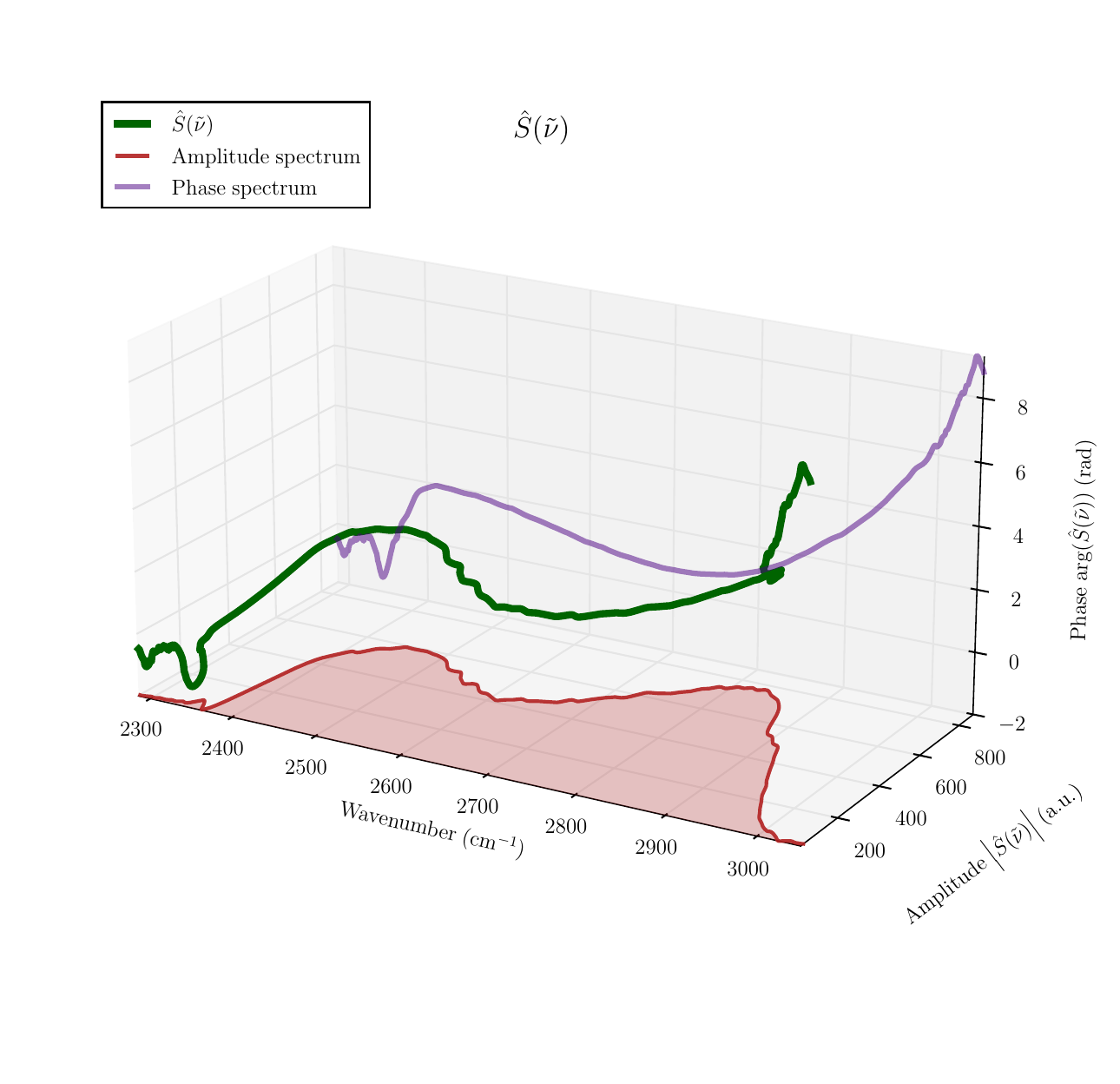}
 \caption{Three-dimensional representation of the QFTIR-retrieved complex spectrum, with amplitude and phase shown as projections; the phase encodes the uncompensated nonlinear dispersion.}
\label{fig:FT-output}
\end{figure}

According to the Wiener–Khinchin theorem, the time-domain correlation function and the frequency-domain spectrum form a Fourier-transform pair; in the general cross-correlation case, the resulting spectrum is complex and thus contains both amplitude and phase. However, in contrast to classical interferometry (mutual coherence, i.e., cross correlation of electric fields), the interference in SU(1,1) interferometry can be interpreted as correlation of indistinguishable pair-creation probability amplitudes in the first and second parametric interactions.
Thus, the complex Fourier transform of the time-domain interferogram yields the complex spectrum, from which the spectral amplitude and phase can be obtained:
\begin{equation}
\hat{S}(\tilde{\nu})=\mathscr{F}\{\mathrm{I_{td}}(\delta)\}
= \int_{-\infty}^{\infty} \mathrm{I_{td}}(\delta)\, e^{-i2\pi \tilde{\nu} \delta}\, d\delta,
\end{equation}
where $\delta = 2 \Delta z$ represents the scanning of the reference arm (optical path difference, assumed $n=1$). 
According to Eq.~\ref{phi-g}, the scanning of the reference mirror (simultaneous scanning of the pump and signal phases) scales the same as the classical scanning of the idler (mid-IR) phase~\cite{Gattinger:25}, i.e. $\varphi_i=\varphi_p-\varphi_s$, so $\Delta z = \Delta\varphi/8\pi n\tilde{\nu}$ as expressed in wavenumbers.

The complex spectrum (see Fig.~\ref{fig:FT-output}) is then defined as: 
\begin{equation}
\hat{S}(\tilde{\nu}) = \left|\hat{S}(\tilde{\nu})\right| e^{i\theta(\tilde{\nu})},
\end{equation}

where $\left|\hat{S}(\tilde{\nu})\right|$ is the amplitude reflectivity, not of particular interest for the dispersion compensation (as opposed to QFTIR absorption spectroscopy), but to be used to frame the SPDC band and discard frequencies with zero amplitude components (phase spectra are meaningful only at frequencies with non-zero amplitudes).
The phase spectrum, which essentially manifests the frequency-dependent phase delays expressed in the idler domain, can be extracted as an argument of the complex-valued spectrum $\hat{S}(\tilde{\nu})$:
\begin{equation}
\theta(\tilde{\nu}) = \arg\!\big(\hat{S}(\tilde{\nu})\big)
= \arctan\!\left[\frac{\Im\{\hat{S}(\tilde{\nu})\}}{\Re\{\hat{S}(\tilde{\nu})\}}\right].
\end{equation}

Figure~\ref{fig:FT-output} displays the complex spectrum after the Fourier transform that contains both the amplitude-coded power spectral density $f(\omega_i)$ and the frequency-dependent phase delay.
Thus, the phase $\theta(\tilde{\nu})$ represents the empirical uncompensated dispersion term and can be used for precise dispersion compensation.

The OCT modality employs a grating spectrometer to sample real-valued spectral interferograms $\mathrm{I_{fd}}$, measured in $\lambda_s$ but remapped to the $\tilde{\nu}$-space.
The phase $\theta(\tilde{\nu})$ thus can be exploited for dispersion compensation in frequency-domain OCT imaging. An OCT A-scan (reflectivity profile) is then defined by the inverse Fourier transform with the corresponding QFTIR phase injected: 
\begin{equation}
R(z)=\int_{-\infty}^{\infty} \mathrm{I_{fd}}(\tilde{\nu})\, e^{-i\theta(\tilde{\nu})}\, e^{i4\pi \tilde{\nu} z}\, d\tilde{\nu}
\end{equation}

It should be noted that the sampling frequencies of the QFTIR and OCT modalities and, therefore, of the spectra do not match, since both modes use different measurement mechanics\textemdash both defined by the Nyquist–Shannon-Kotelnikov sampling theorem but in different domains. 
Hence, the QFTIR phase should be numerically framed to the band of the spectral-domain OCT interferograms and re-interpolated to the sample size of the corresponding OCT signals.
In addition, to suppress artifacts arising from measurement noise, particularly in spectral regions with weak SPDC emission, a polynomial fit was used to obtain a smooth phase component.

\begin{figure}[hbt]
    \newcommand{\commonheight}{8.5cm} 
    \begin{subfigure}[b]{0.6\linewidth}
        \begin{tikzpicture}
            \node[anchor=south west, inner sep=0] (img) at (0,0) {%
                \includegraphics[
                    trim=0mm 0mm 0mm 0mm,
                    clip,
                    height=\commonheight
                ]{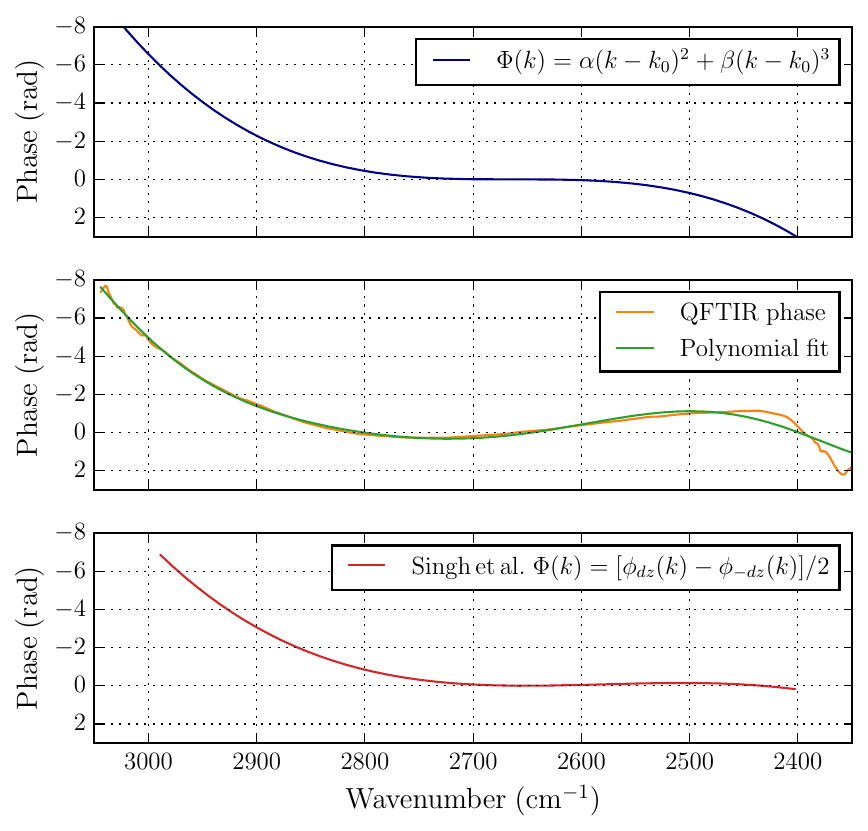}%
            };
            \begin{scope}[x={(img.south east)}, y={(img.north west)}]
            \end{scope}
        \end{tikzpicture}
        \caption{Dispersion phase spectra of an inherently unbalanced nonlinear interferometer retrieved using three different methods}\label{fig:phases}
    \end{subfigure}
    \begin{subfigure}[b]{0.3\linewidth}
        \includegraphics[height=\commonheight]{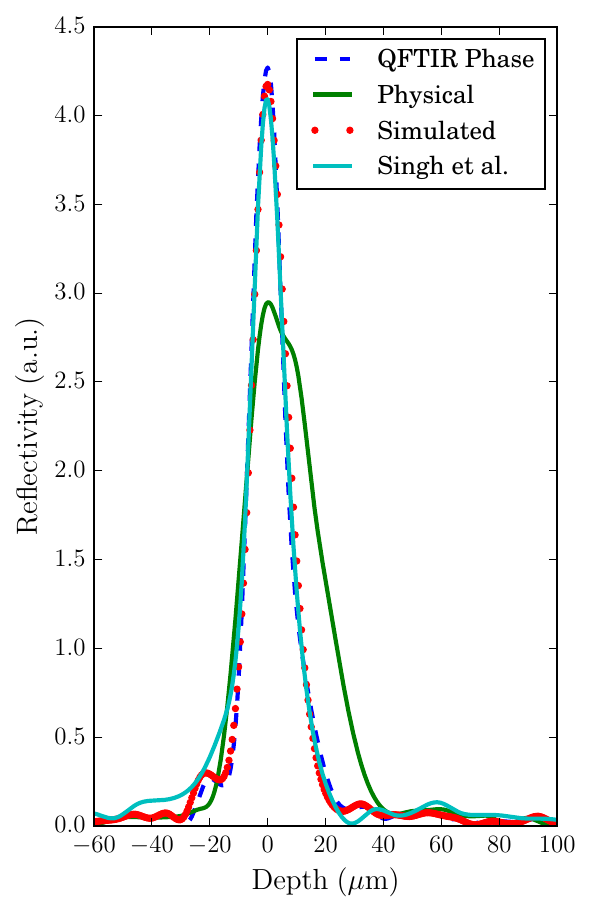}
        \caption{Axial point spread functions for compensation methods}
        \label{fig:psf-and-co}
    \end{subfigure}
    \caption{Derived phase components and results of numerical compensation (demonstrated for corresponding axial point-spread-functions) of the net bi-photon nonlinear dispersion exemplified for different methods; physical compensation is the initial baseline provided by the doublet used in the idler arm.}
    \label{fig:correction}
\end{figure}

The derived phase (QFTIR phase) and the corresponding fit are shown in Fig.~\ref{fig:phases}. In addition, Fig.~\ref{fig:phases} depicts phase terms obtained using two alternative approaches: (i) a fully numerical simulation including second- (GVD) and third-order dispersion terms (modeled as a polynomial by optimizing image sharpness, see Fig.~\ref{fig:correction}), and (ii) an experimentally measured phase spectrum obtained using a method reported in~\cite{Singh_2018}.

The second experimental approach is based on the measurements of spectral interferograms at equal but opposite offsets $\pm dz$ ($\pm$ 300~\textmu m) from the zero delay plane. The spectral phases are extracted using a Hilbert transform, unwrapped, and then subtracted to cancel the carrier term; as a result, the dispersion phase can be obtained directly without performing complex computations or additional measurement procedures. The obtained curves are in good agreement.

The corresponding axial point-spread functions enhanced using the three different methods are shown in Fig.~\ref{fig:psf-and-co}. The physical compensation corresponds to the baseline (no numerical correction) case, in which only the achromatic doublet in the idler arm provides partial dispersion compensation, as described previously. 

It can be observed that, despite the, in principle, very similar results and comparable compensation, the strategy employing injection of the QFTIR phase demonstrates slightly better performance. The difference can be attributed to the high-precision reference interferometric scheme of the QFTIR, which ensures accurate determination of spectral phases, whereas the alternative method relies on manual mirror positioning. The resolution has been improved by a factor of 2.2, from approx. 29~\textmu m to 13~\textmu m (at full width at half maximum). It is should be noted that the proposed calibration-based correction is not limited to the quadratic phase since the measured spectral phase contains high-order terms.

\begin{figure}[hb]
 \centering
        \includegraphics[width=1\textwidth]{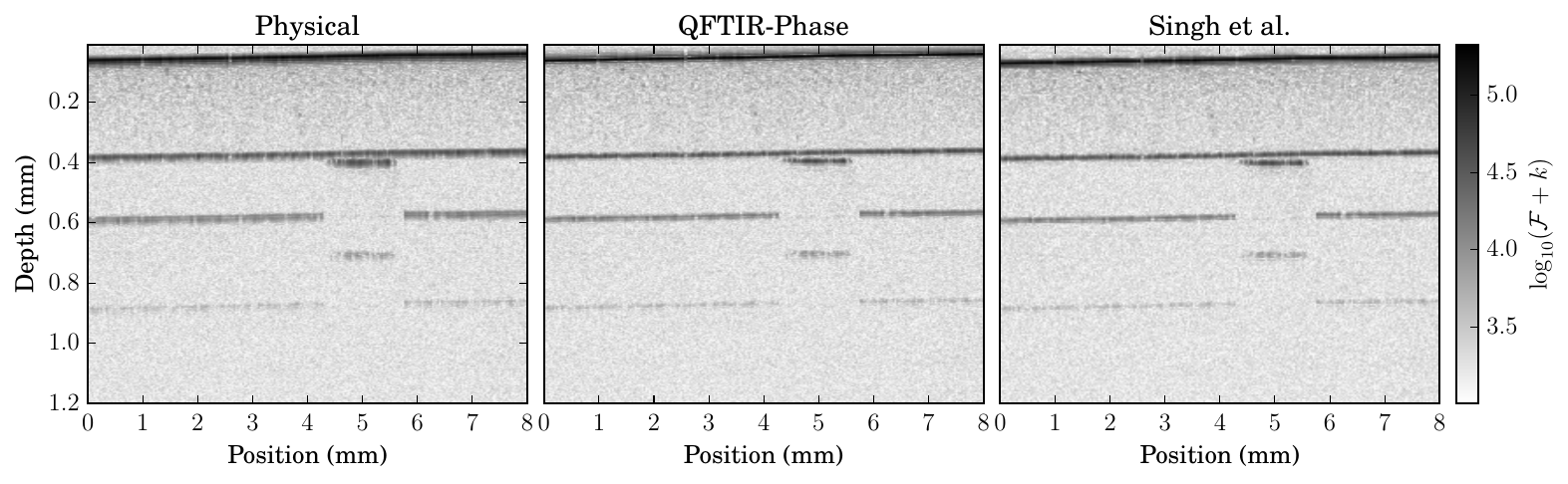}
 \caption{B-scans of sintered alumina ceramics, inaccessible to standard near-IR OCT systems due to the strong scattering, illustrate the proposed method of numerical dispersion compensation using the QFTIR modality of the nonlinear interferometer; the method is benchmarked against the alternative measurement-based compensation techniques.}
\label{fig:bscans}
\end{figure}

Figure~\ref{fig:bscans} shows B-scans of high-scattering alumina ceramics plates (inaccessible to state-of-the-art near-IR OCT~\cite{Su:14}) obtained using two different measurement-based methods for numerical dispersion compensation, alongside the uncompensated scenario (for illustrative purposes).



These B-scans also illustrate the advantages of OCT with undetected photons for non-destructive testing, enabling mid-IR OCT using high-sensitivity, cost-effective near-IR components. A particular advantage that is of great interest in certain areas of application is the ultra-low probing power of only 60~pW, which guarantees that the sample remains unaltered.

\section{Conclusion and outlook}
In this work, we have investigated higher-order dispersion in nonlinear interferometry and assessed its implications for optical coherence tomography (OCT) imaging with undetected photons, an emerging and promising technique for applied non-destructive testing. Moreover, we proposed a numerical compensation method based on measuring the phase term and demonstrated its performance.

Therefore, the distinctive characteristics of bi-photon interferometers, in contrast to classical systems, are presented and discussed. We demonstrated that the energy entanglement of the generated photons (generated in non-degenerate parametric down-conversion) directly results in an effective group-delay dispersion that is the cumulative (additive) contribution of dispersions across the signal and idler spectral bands. 
This has important implications for OCT imaging with nonlinear interferometers and distinguishes the approaches to physical dispersion compensation: instead of matching the dispersion within the interferometric arms (as performed in classical interferometry), bi-photon interferometry requires compensation of the overall (net) dispersion.
In fact, the arms of a Michelson-type nonlinear interferometer only resemble those of a traditional arrangement. The signal and idler photons are not cross-correlated in the classical interferometric way; instead, their mutual emission probability amplitudes corresponding to the first and second passes through the nonlinear crystal are correlated. 
Since a substantial fraction of the uncompensated dispersion originates in the source itself, specifically due to the propagation through the nonlinear crystal, several interesting scenarios arise. In particular, two crystals widely used for short-wave mid-IR metrology with undetected photons, potassium titanyl phosphate (KTP) and lithium niobate (LN), exhibit a zero-dispersion wavelength between the signal and idler bands (broadband systems designed for group-velocity-matching conditions~\cite{Vanselow:19}), resulting in a strong negative net dispersion.
Hence, this can lead to counterintuitive cases (relative to classical experiments) in which adding a positively dispersive medium to either interferometer arm improves the axial resolution. In a particular configuration, a dispersive material can be inserted into the combined beam path after the crystal. These observations emphasize the need for careful engineering of such non-classical OCT systems, since materials and optical components can be deliberately chosen to reduce the effective dispersion. For example, the substrate of a dichroic mirror in the arrangement demonstrated in the experimental section of the manuscript can be replaced with silicon, which exhibits strongly positive dispersion at the idler wavelengths. Alternatively, if these effects are neglected, a pronounced degradation of the axial point-spread function might be expected. Beyond that, it is worth mentioning that Roeder et al. demonstrated engineered, physical GVD cancellation in waveguides (as opposed to bulk crystals considered in this work)~\cite{Roeder_2024}. This approach could provide a path toward on-chip SPDC sources for nonlinear interferometry and OCT with extreme bandwidths and reduced source-induced high-order dispersion.

In the experimental part, we demonstrated the practical impact of dispersion on OCT with undetected photons. We proposed a compensation method that exploits the built-in quantum Fourier-transform infrared (QFTIR) spectroscopic modality to retrieve the nonlinear phase. The method is based on the measurement of time-domain interferograms acquired by scanning the joint signal–pump arm length; the phase component is then extracted using a Fourier transform as an argument of the complex function. Due to the interferometric linearization of the time-domain signal, our approach to dispersion compensation demonstrated enhanced performance; we compared this approach with an established method that similarly relies on direct measurement. Phase terms obtained using alternative retrieval methods were also introduced and discussed. It should be noted that this dispersion-compensation method can be readily transferred to quantitative analysis and dispersion compensation in classical OCT.

\section*{Acknowledgment}
The authors are grateful to Bettina Heise for valuable discussions on dispersion in OCT and for reviewing the derivations to ensure mathematical rigor.

\noindent The authors acknowledge Maryam Viqar, Violeta Madjarova, and Elena Stoykova from Institute of Optical Materials and Technologies Bulgarian Academy of Sciences Sofia for assistance in performing measurements.

\noindent The authors declare no conflicts of interest.

\section*{Funding}
{\"O}sterreichische Forschungsf{\"o}rderungsgesellschaft (FFG), QMIRACT Project, 929209; This project was co-financed by research subsidies granted by the government of Upper Austria: Quick (Wi-2022-597365/18-Au).


\section*{Data availability}
Data underlying the results presented in this paper are not publicly available at this time but may be obtained from the authors upon reasonable request.

\bibliographystyle{unsrtnat}
\bibliography{references}

\end{document}